# Accelerating single-crystal growth by stimulated and self-guided channeling

Yan Ren[1], Changfeng Fang[2], Chengjie Zhu[3], Yvonne Y. Li[4],

Bo Durbeej[5], Xian Zhao[2*], Lu Deng[2,3*]

[1]State Key Laboratory for Crystal Materials, Shandong University, Jinan, China.

[2]Center for Optics Research and Engineering (CORE), Shandong University, Qingdao, China.

[3]MOE Key Laboratory for Advanced Micro-Structured Materials, School of Physics Science and Engineering, Tongji University, Shanghai, China.

[4]Department of Medical Oncology, Dana-Farber Cancer Institute, Harvard Medical School, Boston, MA, USA.

[5]Division of Theoretical Chemistry, Linköping University, SE-581 83 Linköping, Sweden.

*Correspondence: lu.deng@email.sdu.edu.cn (L.D.); zhaoxian@sdu.edu.cn (X.Z)

**Abstract:**

We report a self-guided and "stimulated" single-crystal growth acceleration effect in *static* super-saturated aqueous solutions, producing inorganic ($KH_2PO_4$) and organic (tetraphenyl-phosphonium-family) nonlinear optical single-crystals with novel morphologies. The extraordinarily fast unidirectional growth in the presence of complete lateral growth suppression defies all current impurity, defect and dislocation based crystal growth inhibition mechanisms. We propose a self-channeling-stimulated accelerated growth theory that can satisfactorily explain all experimental results. Using molecular dynamics analysis and a modified two-component crystal growth model that includes microscopic surface molecular selectivity we show the lateral growth arrest is the combined result of the self-channeling and a self-shielding effect. These single-crystals exhibit remarkable mechanical flexibility in winding and twisting, demonstrating their unique advantages for chip-size quantum and biomedical applications, as well as for production of high-yield/high-potency pharmaceutical materials.

**One Sentence Summary:** Discovery of a robust, self-guided and self-stimulated material growth acceleration effect in aqueous inorganic and organic single-crystals.

Single-crystal (*1,2*) is a catalyst for diverse scientific advances in optical science (*3,4*), chemical and biomedical science (*5-7*), mineralogy (*8*), interstellar astro-chemistry (*9,10*), and even pharmaceutical (*11-14*) development. It has significantly impacted many scientific fields ranging from quantum entanglement and communication technologies (*3,4*) to chemical engineering, protein crystallization (*13*) and pharmaceutical manufacturing of high-potency single-crystal/co-crystal drugs (*12,14*). For decades, self-restricted growth of certain crystal faces, even in a supersaturated solution that favors growth, has been a subject of active investigations. The well-established Cabrera-Vermilyea theory (*15*) on surface chemistry by impurities has stimulated extensive research and it is now widely accepted (*16*) that crystal growth eventually arrests by one of two means: (i) the depletion of growth molecules from the surrounding solution when the super-saturation reaches zero, or (ii) the presence of metal impurities that "poison" surface chemistry, thereby inhibiting the crystalline growth of native molecules (*15-24*). Here, we report a single-crystal growth phenomenon that defies this well-established theoretical understanding. Indeed, the observed growth bi-morphology cannot be explained by any currently-known theories of crystalline surface growth inhibition mechanisms, including all impurity, restricted-nucleation, dislocation or defect-based theories, as well as solid-state chemistry and material science in general, a strong indication of unknown surface physics effects and processes. We propose a self-channeling stimulated matter growth acceleration theory, in analogy to the coherent stimulated Raman light scattering process in optical science, and show that it can satisfactorily explain all experimental results.

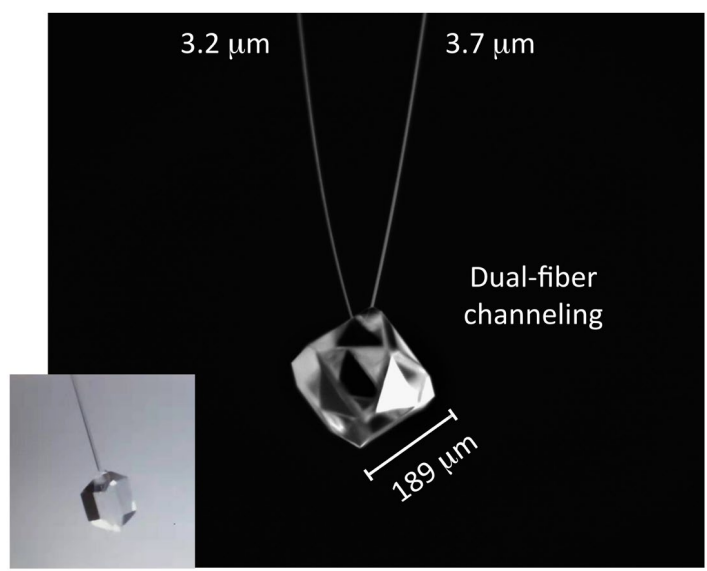

Fig. 1 **Bi-morphologies of single-crystalline pendants grown by self-channeling stimulated growth acceleration technique**. Self-channeling stimulated growth of a bulk crystal fused to two single-crystal fibers that are under lateral growth arrest. Inset: Single-crystal KDP fiber with its pyramidal face fused to the pyramidal face of a bulk crystal of the similar size (see Fig. 4E for the corresponding simulation). The growth of the fiber is completely suppressed whereas the bulk crystal grows un-impeded to nearly ninety times larger in a few minutes.

Figure 1 shows bi-morphologies of inorganic KH$_2$PO$_4$ (KDP) single-crystals grown by the novel stimulated single-crystal growth acceleration technique. Similar bi-morphologies using organic TTP-X family single crystals (X: Cl, Br, I, etc.) have also been produced. To achieve this, we bring the ends (also referred to as pyramidal faces) of two single-crystal fibers (aspect ratio $R \sim 60$, radius $\sim 3$ μm, length $\sim 180$ μm) and a small bulk single-crystal (initial size $\sim 3$ μm, $R \sim 2$) in contact *in situ* in an aqueous solution. In the subsequent growth period we observed no growth on the fibers' lateral faces (also referred to as prismatic faces). However, the small bulk crystal grows with an astonishingly fast rate of $v > 350$ nm/s [see (*26*) for more] in *each direction* (from $\sim 3$ μm initial size to nearly 190 μm in just a few minutes!). That is, in the *same growth environment and proximity* the fiber sections experience complete prismatic face growth arrest and yet the bulk crystal grows steadily unimpeded to more than sixty times of its original size in just a few minutes. When the pyramidal face of a single-crystal fiber is fused to the pyramidal face of a small bulk single-crystal (Fig. 1 inset) the latter can grow to nearly ninety times larger in a few minutes during which the fiber section does not grow at all. This bi-morphology clearly demonstrates that impurity-based growth inhibition mechanisms do not play any important role. Indeed, **there is no reasonable explanation as to why impurities *in the same solution and proximity* only severely impact and suppress the lateral growth characteristics of the fiber section, while leaving the growth dynamics of the bulk crystal completely intact.** Furthermore, extensive X-ray diffraction studies show extremely sharp single peaked spectra that rule out any defect/dislocation-based arguments [see Supplementary Material (SM) I]. This bi-morphology rapid growth of single crystals is a testimony to the robust self-channeling-stimulated matter growth acceleration mechanism we explain below.

Seed crystals with large aspect ratios can produce a solute self-channeling effect similar to the "point-effect" in electrostatics. This effect makes the pyramidal face a strong "solute collector", creating a steep concentration gradient at the pyramidal face. This solute gradient results in a pyramidal-face-guided solute flow that enables an accelerated matter growth phenomenon never before observed in the field of crystallography and material science. Figures 2A and 2B show the pyramidal face growth rates as functions of $R$ for KDP and TPPCl crystals subjecting to strong prismatic face growth arrest. Remarkably, the measured growth rate can be well-approximated by $v \propto (\ln[C_\downarrow(R)])^n$ where we introduce an "$R$-dispersive" *local* concentration $C_\downarrow(R)$ in analogous to the resonant/near-resonant frequency dispersion of molecular optical polarizability (*25*). The crystal growth driving force (GDF) for such a highly dynamic "$R$-dispersive" process can be expressed as [see SM II, III]

$$GDF \propto \ln\left(\left|d_\downarrow^{(1)} \cdot \nabla C_\downarrow(R)\right|\right), \tag{1}$$

in analogous to the coherent stimulated Raman light scattering enhancement by lattice normal mode vibrations (*25*). The physical meaning of the tensor $d_\downarrow^{(1)}$ is the strength of "*R-sloped*" contribution, i.e., the *R*-dependent growth acceleration, arising from the *microscopic* concentration gradient. Therefore, the postulation of "*R-dispersive*" *local* concentration and Eq. (1) collectively represent the concept of "Raman-like" stimulated material growth mechanism. Indeed, the excellent fits in Figs. 2A and 2B to Eq. (1) clearly demonstrate that it is the local concentration *gradient* near the pyramidal face, *i.e.*, $\nabla C_\downarrow$, rather than the quasi-steady macroscopic solution concentration as conventional theories have always assumed, that drives the accelerated surface growth. For sufficiently large *R* Figs. 2A and 2B indicate the presence of a characteristic aspect ratio that acts as an effective "healing length" beyond which the self-channeling effect saturates.

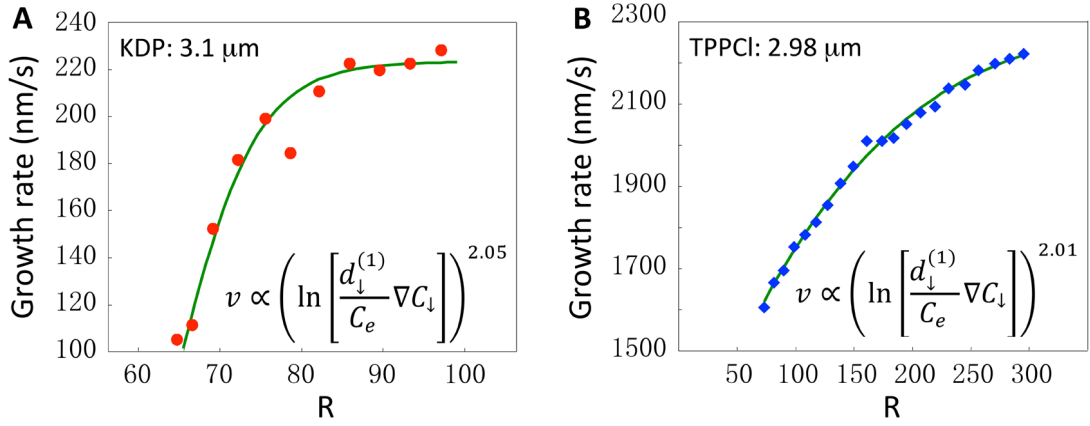

Fig. 2 **Self-channeling-stimulated growth acceleration effect.** Pyramidal face growth rate as a function of fiber aspect ratio *R* for KDP (**A**) and TPPCl (**B**) single-crystalline fibers subjecting to prismatic face growth suppression. Green curves are fits using Eq. (1).

The idea of stimulated matter growth mechanism that defies all currently accepted theories arises from a series of single-crystalline fiber studies. We prepare and submerge single-crystalline KDP/TPPCl/TPPBr fibers (typically diameter ~ 3 μm, length 50 ~ 100 μm) in ultra-high purity supersaturated (σ = 0.237) aqueous solution at room temperatures [see SM I]. The growth process is *static* and we routinely obtain single-crystalline KDP/TPPCl/TPPBr fibers with large aspect ratio *R* > 1000 [Figs. 3A, 3B], excellent optical properties and remarkable mechanical flexibility in twisting [Fig. 3C] and winding [Fig. 3D].

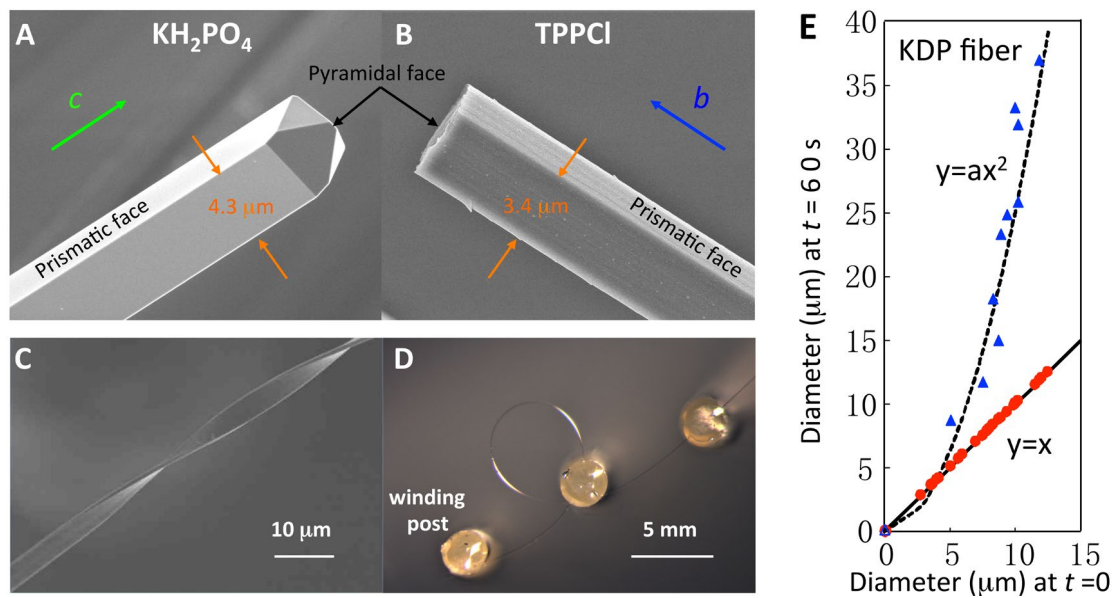

Fig. 3 **Images of single-crystalline organic and inorganic nonlinear crystal fibers.** Both KDP (**A**) and TPPCl (**B**) show prismatic face growth suppression (diameter remain unchanged) whereas the growth of pyramidal faces is unimpeded. They exhibit remarkable mechanical flexibility for twisting (**C**) and winding (**D**) (winding posts are optical fibers). (**E**) Prismatic face growth suppression (red dots) and quadratic growth characteristics of fibers with imperfections (blue triangles).

From a single-crystal growth perspective, the most striking feature shown in Figs. 3A and 3B is the rapid pyramidal face growth coupled with complete prismatic face growth suppression. We observed no change in prismatic faces in a 15-hour growth period during which the crystal pyramidal face grows at a fast rate (> 400 nm/s in a *static* super-saturated solution for KDP and even much faster for TPP-X family single crystals), yielding crystal fibers with aspect ratios $R > 1000$. This is in stark contrast to the growth of a low-$R$ KDP/TPPCl/TPPBr crystal under the same growth conditions where *all* surfaces grow with a similar slower rate [< 30 nm/s (*26*)], resulting in a bulk crystal typically having $R < 3$. Figure 3E shows the diameter of a growing fiber as a function of its initial diameter in a fixed 60 seconds period. Fibers with excellent uniformity (verified by microscope inspection) follow a perfect $y = x$ relation, indicating complete lateral growth suppression ($x$ and $y$ are crystal-fiber diameters measured at $t=0$ and $t=60$ seconds). However, fibers with non-uniformity (by microscope inspection) exhibit a quadratic dependency on their initial seed diameters. We speculate that surface deformation may result in local field irregularities that inhibit the formation of an effective shielding layer (see discussion later), therefore allows the growth process to reconstitute.

Molecular adsorption is the first critical step in surface growth processes. In general, solute transport by macroscopic diffusion is insensitive to the orientation of individual molecules (neglecting molecular geometry). This property changes when molecules reach the crystallization surface where molecular "orientation" and surface interaction potential introduce surface-chemistry-based molecular orientation selectivity (*27,28*). As a result, molecules with a favorable surface-heading orientation are energetically predisposed for surface adsorption, whereas molecules with unfavorable orientations have low surface bonding probabilities. For sufficiently high concentration the non-adsorbed molecules will accumulate near the surface, forming an "inert layer" that blocks the penetration of arriving molecules with favorable orientations. As the diffusion process continues, driven by the *macroscopic* concentration gradient, this shielding layer increases and surface growth along this dimension quickly stops. We note that the pyramidal-face-guided flow arising from the steep local concentration gradient can also impact the growth dynamics of prismatic faces. It drains solute from the region above the prismatic face, depleting arriving molecules and therefore enhancing the prismatic face growth suppression effect. For simplicity and without the loss of generality, we use KDP molecules in following discussions although we have also carried out detailed analyses for ions, unit cells and clusters [see SM IV].

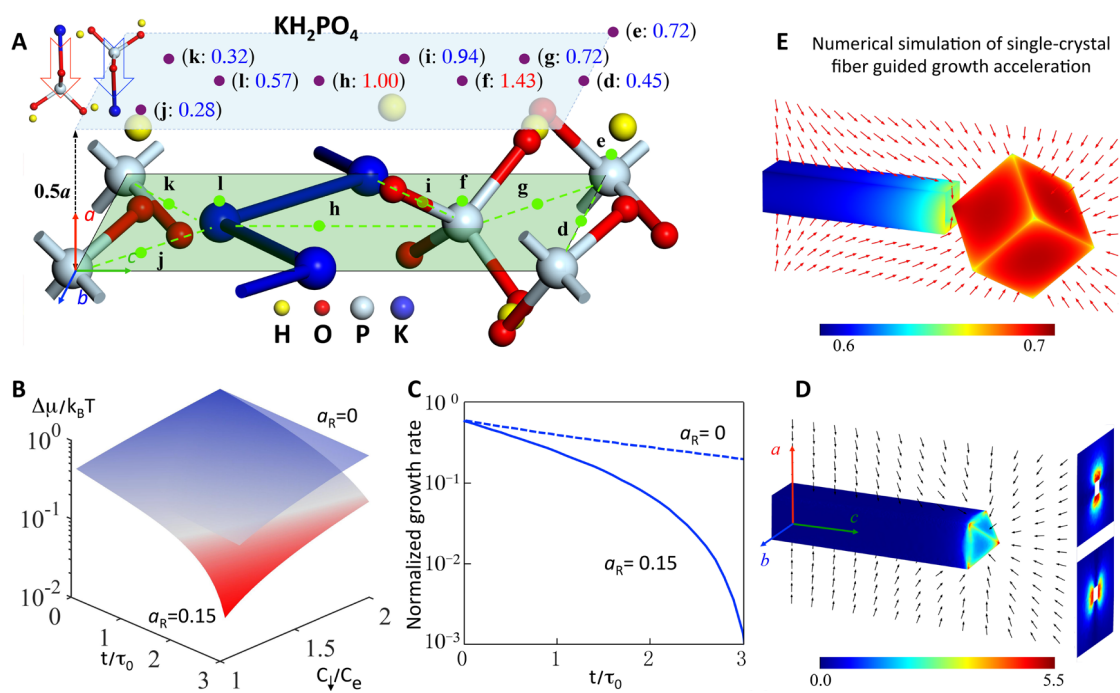

Fig. 4 **Molecular-orientation-dependent surface energy, crystal driving force and growth rate, microscopic concentration gradient.** (**A**) Surface energy ratios at *0.5a* ($a$ = 7.45280 Å) above the KDP (100) prismatic surface for different surface sites and molecular headings orientations. (**B** and **C**) Dimensionless crystal growth driving force GDF = $\Delta\mu/k_BT$ and growth rate as functions of $C_\downarrow/C_e$ and normalized growth time for growth impedance factor $a_R = 0$ (top flat surface and dashed-line) and $a_R = 0.15$. (**D**) Local concentration gradient $|\hat{z} \cdot \nabla C_\downarrow|$ near the prismatic surface showing pyramidal-face-guided surface flow (shown as arrows). Contour plots on the right show the gradient distribution in the x and y directions near the pyramidal face. (**E**) Numerical simulation of accelerated growth of a small bulk crystal fused to a fiber with large *R* (see Fig. 1). Here, the red color indicates low concentration gradient.

We carried out extensive numerical simulations in which surface adsorption energy for nine sites on a KDP (100) face with different molecular surface heading configurations are evaluated using density functional theory methods (*29-31*). Figure 4A reveals that the K-heading orientation is the energetically dominant surface adsorption channel. Here, numbers in parenthesis indicate the ratio of surface adsorption energies $E_{PO4\text{-heading}}/E_{K\text{-heading}}$. The less the blue number than 1 the greater the K-heading dominance. Likewise, the greater the red number than 1 the greater the PO$_4$-heading dominance. Calculations for distances between $0.185a$ to $0.5a$ above KDP (100) face using molecules, ions, unit-cells, as well as clusters as crystallization elements show the similar surface molecular selectivity tendency.

The surface molecular-orientation selectivity, and its dynamic impact to surface crystallization, can also be explained using an intuitive two-component diffusion model without external agitation [SM II-IV]. We note that while in reality molecules have a distribution of orientations it is the quasi-binary property of the crystalline surface that breaks the orientation distribution symmetry and gives rise to surface bonding selectivity. We emphasize that this two-component model is based on the well-known crystal growth theory (*32*) but with added *microscopic surface reaction selectivity and dynamics*. We introduce a crystal surface growth impedance parameter $a_R$ to characterize the impact on the bonding probability of molecules with favorable heading-orientation by the accumulation of molecules with unfavorable orientation. In fact, *when only one molecular component is considered this two-component model reduces exactly to the well-known result* (*32*) *of crystal growth, as it must*.

Figures 4B and 4C show normalized $GDF = \Delta\mu/k_B T$ and growth rate as functions of shielding parameter $a_R$ [SM II-IV]. A non-vanishing $a_R$ leads to rapid reduction of $C_\downarrow$ on the prismatic face. Consequently, both GDF and crystallization rate quickly approach zero and the growth of the prismatic face ceases. Figure 4D shows the color contour plot of the numerical solution of the 3-dimensional diffusion equation. It reveals that the large concentration gradient leads to strong local solute flow, resulting in rapid growth of the pyramidal face that prevents the formation of *a local "shielding layer"* near the pyramidal face.

The self-channeling-stimulated growth acceleration mechanisms can act robustly, independently and simultaneously on multiple seed crystals with more complex arrangement, as shown in Fig. 1 (experiment) and Fig. 4E (numerical simulation). This opens a new frontier in modern optical engineering with the possibility of independent growth manipulation for complex single-crystal-based photonic architectures. The demonstrated flexibility for twisting and winding makes the new technique very attractive for nonlinear guided-wave devices. One immediate ramification is the impact to the chip-based coherent light generation, via the large Raman-shift of mm-sized TPPX family single crystals, in wavelength regions currently not accessible without employing bulky high-energy pulsed laser systems

(see SM V). In addition to these optical engineering applications, the accelerated single-crystal growth technique may also broadly impact engineering of future high-yield/high-potency crystal drugs, crystallized proteins and biomedical materials. We finally note that the self-channeling-stimulated growth acceleration theory explains all experimental results reported rather well. This further illustrates the rich physics of crystallization processes in different growth configurations and environment, even for a system that has been thought to be well understood.

# References and Notes

1. D. W. H. Rankin, CRC handbook of chemistry and physics, 89th edition, edited by David R. Lide. *Crystallogr. Rev.* **15**, 223-224 (2009).
2. R. A. Laudise, *The growth of single crystals*. (Prentice Hall, 1970).
3. A. Yariv, P. Yeh, *Optical waves in crystals*. (Wiley, 1984), vol. 5.
4. S. Olmschenk, Linking crystals with a single photon. *Nat. Photonics* **6**, 221-222 (2012).
5. A. J. Malkin, Y. G. Kuznetsov, T. A. Land, J. J. DeYoreo, A. McPherson, Mechanisms of growth for protein and virus crystals. *Nat. Struct. Biol.* **2**, 956-959 (1995).
6. D. E. McRee, *Practical protein crystallography*. (Elsevier, 1999).
7. A. McPherson, *Crystallization of biological macromolecules*. (Cold Spring Harbor Laboratory Press, 1999).
8. C. Burns Peter, J. Finch Robert, Wyartite: Crystallographic evidence for the first pentavalent-uranium mineral. *Am. Mineral.* **84**, 1456-1460 (1999).
9. C. F. Boutron, *Topics in atmospheric and interstellar physics and chemistry*. (EDP Sciences, 1994), vol. 1.
10. E. Krugel, *An introduction to the physics of interstellar dust*. (CRC Press, 2007).
11. M. Daniels, Why crystals could be the shape of future pharmaceuticals, <https://phys.org/news/2015-06-crystals-future-pharmaceuticals.html> (June 8, 2015).
12. N. Chaisongkram, S. Maosoongnern, A. E. Flood, Unusual Crystal Growth Kinetics of (RS)-Ibuprofen from Ethanolic Solutions. *Chem. Eng. Technol.* **42**, 1519-1524 (2019).
13. B. Y. Shekunov, P. York, Crystallization processes in pharmaceutical technology and drug delivery design. *J. Cryst. Growth* **211**, 122-136 (2000).
14. Z. Gao, S. Rohani, J. Gong, J. Wang, Recent Developments in the Crystallization Process: Toward the Pharmaceutical Industry. *Engineering* **3**, 343-353 (2017).
15. R. H. Doremus, B. W. Roberts, D. Turnbull, *Growth and perfection of crystals: proceedings*. (Wiley, 1958).
16. M. Sleutel, J. F. Lutsko, D. Maes, A. E. S. Van Driessche, Mesoscopic Impurities Expose a Nucleation-Limited Regime of Crystal Growth. *Phys. Rev. Lett.* **114**, 245501 (2015).
17. S. Y. Potapenko, Moving of step through impurity fence. *J. Cryst. Growth* **133**, 147-154 (1993).
18. L. N. Rashkovich, N. V. Kronsky, Influence of $Fe^{3+}$ and $Al^{3+}$ ions on the kinetics of steps on the {1 0 0} faces of KDP. *J. Cryst. Growth* **182**, 434-441 (1997).
19. K. Sangwal, Effects of impurities on crystal growth processes. *Prog. Cryst. Growth Charact. Mater.* **32**, 3-43 (1996).
20. B. Lewis, The growth of crystals of low supersaturation: I. Theory. *J. Cryst. Growth* **21**, 29-39 (1974).
21. M. Bohenek, A. S. Myerson, W. M. Sun, Thermodynamics, cluster formation and crystal growth in highly supersaturated solutions of KDP, ADP and TGS. *J. Cryst. Growth* **179**, 213-225 (1997).
22. W. J. P. van Enckevort, A. C. J. F. van den Berg, Impurity blocking of crystal growth: a Monte Carlo study. *J. Cryst. Growth* **183**, 441-455 (1998).
23. N. Zaitseva, L. Carman, I. Smolsky, R. Torres, M. Yan, The effect of impurities and supersaturation on the rapid growth of KDP crystals. *J. Cryst. Growth* **204**, 512-524 (1999).

24. T. A. Land, T. L. Martin, S. Potapenko, G. T. Palmore, J. J. De Yoreo, Recovery of surfaces from impurity poisoning during crystal growth. *Nature* **399**, 442-445 (1999).
25. Y. R. Shen, N. Bloembergen, Theory of Stimulated Brillouin and Raman Scattering. *Phys. Rev.* **137**, A1787-A1805 (1965).
26. The comparison is made with respect to a small bulk crystal at rest. See discussons at the end of Supplementary Material II.
27. H. Hosono, S. Kawamura, Y. Abe, Prefered molecular orientation at crystal growth front in Ca(PO3)2 glass. *J. Mater. Sci. Lett.* **4**, 244-246 (1985).
28. S. Whitelam, Y. R. Dahal, J. D. Schmit, Minimal physical requirements for crystal growth self-poisoning. *J. Chem. Phys.* **144**, 064903 (2016).
29. P. Giannozzi *et al.*, QUANTUM ESPRESSO: a modular and open-source software project for quantum simulations of materials. *J. Phys.: Condens. Matter* **21**, 395502 (2009).
30. L. Zhang, Y. Wu, Y. Liu, H. Li, DFT study of single water molecule adsorption on the (100) and (101) surfaces of KH2PO4. *RSC Adv.* **7**, 26170-26178 (2017).
31. C. Fang, C. Zhu, X. Zhao, L. Deng, Strong Mesoscopic Prismatic Face Growth Suppression in High-Speed Unidirectional Growth of KDP Single Crystals. *Cryst. Growth Des.* **20**, 3531-3536 (2020).
32. P. Rudolph, Transport phenomena of crystal growth—heat and mass transfer. *AIP Conf. Proc.* **1270**, 107-132 (2010).

**Acknowledgments:**

YR acknowledges Prof. Chengqian Zhang of the State Key Laboratory of Crystal Materials of Shandong University for technical assistance in crystal growth.

**Funding:** YR was support by National Natural Science Foundation of China (Grant No. 51872163) and Natural Science Foundation of Shandong Province (ZR2019MEM006). CFF was supported by the Key Research and Development Program of the Shandong Province (Grant No. 2018GGX102008) and the Fundamental Research Funds of Shandong University. CJZ acknowledges the supports of the National Key Basic Research Special Foundation (2016YFA0302800); Shanghai Science and Technology Committee (18JC1410900) and National Natural Science Foundation of China (11774262). XZ was supported by the Primary Research & Development Plan of Shandong Province (2017CXGC0413). L. Deng acknowledges support from the Fundamental Research Funds of Shandong University.

**Author contributions:** YR (ry@sdu.edu.cn), CFF (cfang@sdu.edu.cn), CJZ (cjz@tongji.edu.cn), and YYL (yvonney_li@dfci.harvard.edu) contributed equally to this work. LD (lu.deng@email.sdu.edu.cn) conceptualized the project, designed experiments, and developed the self-channeling stimulated growth acceleration theory. LD also developed the two-component molecular-orientation-based self-shielding model and the theoretical interpretations of the gradient-dependent crystal driving force. YR grew crystal samples under the guidance of LD. CFF performed molecular density functional simulations under the guidance of XZ (zhaoxian@sdu.edu.cn) and LD. CJZ performed numerical calculations of diffusion equations and data analysis under the guidance of LD. LD wrote the initial manuscript. YYL and BD (bo.durbeej@liu.se) made key contributions to the final manuscript composition. All authors contributed to the manuscript revisions.

**Competing interests:** Authors declare no competing interests.

**Data and materials availability:** Data in the main text and supplementary materials is available.

**Supplementary Materials:**

Materials and Methods

Supplementary Text

Figs. S1 to S2

References

# Supplementary Materials for

# Accelerating single-crystal growth by stimulated and self-guided channeling

Yan Ren[1], Changfeng Fang[2], Chengjie Zhu[3], Yvonne Y. Li[4],

Bo Durbeej[5], Xian Zhao[2*], Lu Deng[2,3*]

**This PDF file includes:**

    Materials and Methods
    Supplementary Text
    Figs. S1 to S2
    References

**Materials and Methods**

**I. Single-crystalline KDP fiber seed preparation**

We use ultra-high purity KDP solution dedicated for production of research grade KDP crystals for high power laser applications. The purity of the solution at selected super-saturation is strictly and actively controlled so that the metal impurity contamination is limited to a few parts per $10^7$ (usually < 200 ppb). Typically, seed crystals with diameters about 5 μm are grown in 4.99g (solute)/100g ($H_2O$) KDP solution at 298 K (super-saturation 23.76%). The solution temperature is stabilized at 298 $\pm$ 0.1 K and the seed crystals are produced on a substrate after fast cool-down and evaporation at room temperature. Extensive Scanning Electron Microscopy imaging and X-ray diffraction (XRD) analysis are performed to verify that these micro-fibers are KDP single crystals having point symmetry ($\bar{4}2m$) and space group I$\bar{4}$2d, with crystal lattice parameters of $a = b$ = 7.45280 Å, $c$ = 6.97170 Å and $\beta_{ab} = \beta_{ac} = 90°, \beta_{bc} = 90°$, respectively.

The technique for making TPPCl single crystal fibers follows a similar protocol. Detailed description and parameters will be published elsewhere (*1*).

Data reported in this work are obtained under the isothermal conditions given above, although we have also carried out extensive studies on growth dynamics under different temperatures and super-saturations.

The extremely sharp XRD spectra of single-crystalline KDP and TPPCl fibers [Figs. S1A and S1C] in comparison with corresponding powder XRD spectra [Figs. S1B and S1D)] show no any observable spectra line broadening, indicating no detectable dislocations, defects and other structural deformation mechanisms. This conclusion is also verified with microscope inspections of various crystal surfaces. Therefore, dislocation and defect-based surface growth blockages are not relevant mechanisms for the observed prismatic face growth arrest reported in the text. **Physically, there is no any reasonable explanation as why dislocations or surface defects, if exit, can uniformly block the growth of all four prismatic faces for such a long linear dimension while still allow the rapid growth of the pyramidal face.**

**II. Two-component diffusion equations with surface selectivity**

The surface molecular-orientation selectivity, and its dynamic impact to surface crystallization and growth, can be understood using a two-component diffusion model without external force (*2*),

$$\dot{C}_j + \nabla \cdot \left( -D \nabla C_j \right) = 0 \ ; \ (j = \uparrow \text{ or } \downarrow) \tag{SM1}$$

subjected to time-dependent boundary conditions

$$\dot{C}_j = \widetilde{K}[C_0 - (C_\uparrow + C_\downarrow)] + b_j\,[-k_\downarrow(C_\downarrow - C_e) - a_R C_\uparrow]; \quad (b_{\uparrow;\downarrow} = 0; 1). \quad \text{(SM2)}$$

Here, $C_\uparrow$ and $C_\downarrow$ represent concentrations of molecules with unfavorable and favorable orientations near the crystallization surface (also see Sec. IV), respectively, and $C_e$ is the equilibrium concentration on the liquid-crystalline surface. $D$, $\widetilde{K}$ and $k_\downarrow$ are the solute diffusion constant, normalized diffusion time constant, and surface kinetic coefficient of the $C_\downarrow$ component, respectively [for calculation convenience we have normalized Eqs. (SM1) and (SM2) with respect to a characteristic time constant $t_0$ so that the time derivative is with respect to $\tau = t/t_0$]. In Eq. (SM2), we introduce $a_R$ to characterize the impact on the mobility of molecules with favorable orientation by the dynamically evolving shielding layer (i.e., through collision effects etc.). We note that while the first term in Eq. (SM2) describes the macroscopic transport of solute, the last two terms, which describe surface chemistry processes for the $C_\downarrow$ component only, introduce *local* surface properties such as adsorbent orientation selectivity that directly impact the surface crystallization process. ***We emphasize that when only one molecular component is considered Eqs. (SM1) and (SM2) reduce to the well-known result (3) of surface growth by taking $a_R = 0$, as it must.***

Solving Eqs. (SM1) and (SM2) numerically we obtain $C_\downarrow(\tau; \mathbf{r})$ as a function of space and time from which the local super-saturation, crystal growth driving force $\Delta\mu/k_B T = \ln(C_\downarrow/C_e)$ (4) and growth rate can be computed. Here, $\Delta\mu$ is the difference in the chemical potential, $k_B$ is the Boltzmann constant, and $T$ is the absolute temperature. This yields Figs. 4B, 4C and 4E in the text. Furthermore, the solution also gives a full crystal growth dynamics where surface solute flow and local concentration gradient as the driver of accelerated growth can be understood. The color contour shown in Fig. 4E at different position $z$ on the prismatic face depicts the gradient $|\hat{\mathbf{z}} \cdot \nabla C_\downarrow|$. Likewise, $|\hat{\mathbf{x}} \cdot \nabla C_\downarrow|$ and $|\hat{\mathbf{y}} \cdot \nabla C_\downarrow|$ can also be similarly obtained, giving a full description of the concentration gradient near the pyramidal surface. Note that on the prismatic face the concentration gradient increases toward the pyramidal face, indicating a self-guided surface solute flow as the self-channeling theory predicted. Near the center of the pyramidal surface the magnitude of the local concentration gradient increase rapidly, signifying substantially accelerated growth. Parameters used in numerical calculations: initial solution concentration is normalized to unity, $C_\uparrow(0) = C_\downarrow(0) = 2C_e = 0.5$, $\widetilde{K} = 0.5$, $k_\downarrow = 0.25$, $R = 35$, and $a_R = 0.15$, respectively. Other parameter combinations yield similar conclusion.

We have also carried out calculations for a crystal with low $R$ = 3. As expected, due to the lack of strong self-channeling-stimulated effect for small $R$, the surface solute flow along the prismatic face is non-noticeable. *More importantly, the concentration gradient towards the pyramidal face is substantially weak and is nearly isotropic and symmetric, indicating lack of any growth acceleration effect.* As a result, all faces grow slowly in all directions, yielding a bulk crystal with low R as typical bulk crystal growth in static growth environment.

It is worthy of further clarifying the above described slow growth rate of a small bulk crystal with low $R$ (also see relevant discussions in reference to (*26*) in the text). Under the same super-saturated growth conditions the growth rate of a small bulk crystal physically unrestrained is about a factor of 3 slower than that of the fiber pyramidal face. However, this is not the correct growth rate to compare with. This is because due to the high super-saturation a micron size bulk crystal with low $R$ dances around due to the reaction force associated with the crystal growth. This unrestrained situation with substantial momentum-transfer-based motion is equivalent to the crystal growth under strong external agitation. The fiber growth does not show such an agitated motional behavior. This is because that all prismatic surfaces suffer growth arrest, remarkably avoiding any lateral motion by growth force. Furthermore, the strong, guided flow is symmetrically parallel to prismatic faces, resulting in well-balanced steady lateral position even though the pyramidal face grows with a high speed. Therefore, the relevant growth rate for comparison should be made in reference to a small bulk crystal with its motion restrained. Under such a motion-restrained condition the growth rate of a small bulk crystal is much slower, typically a factor of 10 or more slower (< 30 nm/s as mentioned in the text), than that of the pyramidal face growth rate of the fiber. These conclusions are in agreement with fast growth technique reported in literature where crystals are forced to rotate fast in the solution. However, it is very important to note that single crystals grown using such strongly agitated flow techniques (including the above unrestrained micro-crystal growth condition) have significantly reduced optical qualities.

**III. Local concentration-gradient-stimulated growth acceleration**

Local concentration gradient of the orientation-favorable solute and the key role it plays in fast, stimulated crystal growth can be argued and analyzed using the numerical solution of Eqs. (SM1) and (SM2), and in comparison with crystal growth rate measurements.

To understand the stimulated growth acceleration effect using the above model we first calculate concentration $C_\downarrow(R)$ at a fixed point on the pyramidal face by numerical solving two-model diffusion equations Eqs. (SM1) and (SM2). The solution is then fit with a generalized sigmoidal function with high accuracy (Reduced–$\chi^2 \approx 10^{-6}$, Reduced-Square $= 0.99985$, and the largest fitting parameter standard error being $< 8.5 \times 10^{-3}$). In the second step, we fit measured growth rate as a function of $R$ using the crystal growth rate function $v \propto (\ln[F(R)])^n$ derived from the Gibbs-Duhem theory (*4*). Here, $F(R)$ is a generalized sigmoidal function of $R$ with four arbitrary fitting parameters including the exponent *n* (*n* = 2 by quantum mechanical calculations). Excellent fits to data, both for inorganic and organic single-crystalline fibers, are obtained as shown in Figs. 2A and 2B in the text. We find that the functional form of $F(R)$ by fitting to the data agrees well with $\left[ln\left(d_\downarrow^{(1)} \cdot \nabla C_\downarrow(R)/C_e\right)\right]^2$, as predicted by Eq. (1) in the text. Thus, it is the local concentration gradient, i.e., $\nabla C_\downarrow(R)$, rather than the local steady concentration $C_\downarrow(R_p)$, that plays the dominant role in the accelerated growth of the pyramidal face. The steep increase of the rate as $R$ increases signifies an accelerated growth process not known before in crystallography. We note that this self-channeling-stimulated growth acceleration mechanism is very similar to the stimulated Raman light scattering enhancement where the scattering process is *strongly enhanced by the derivative* of local molecular lattice normal mode vibrations (*5*). Therefore, we propose, in analogous to the molecular normal mode vibration based stimulated Raman scattering theory, an "*R-dispersive*" microscopic local concentration function as

$$C_\downarrow(R) = C_\downarrow(R_0) + \sum_m d_\downarrow^{(m)} [\nabla^m C_\downarrow(R)]_{S(R_0)} (R - R_0)^m.$$

Here, $R_0$ is the aspect ratio measured from a reference point to the pyramidal surface $S(R_0)$, and the coefficient tensor $d_\downarrow^{(m)}$ characterizes the strength of the *m*-th order derivative contribution from the *local* concentration. The above postulation and the Eq. (1) in the text represent a new stimulated matter growth theory we propose in this research. This theory can explain all experiment data, which cannot be explained by any of current-known theories, satisfactorily.

**IV. DFT calculations and considerations of other solute components**

Crystalline surface growth is a complex process (*3,4,6-9*) especially in cases of high super-saturation. Surface chemistry, growth energetics, molecules and cluster formation and approaching configurations, group dissociations, ion reattachment, surface dehydrations and bond relaxations and many other mechanisms all can influence and contribute to the crystallization processes in different super-saturation and temperature regimes. In the present work we focus on finding physical environment and crystal geometry induced and stimulated growth effects. The framework discussed in the text appears to be sufficient for explaining the accelerated pyramidal face growth and the prismatic face growth arrest encountered in our experiments. Using the widely practiced molecular dynamics algorithm (*10*) we calculate the KDP (100) surface adsorption energy at different sites and distances above the (100) face for approaching KDP molecules with different heading orientations. This how the surface adsorption energy ratios for different heading orientations shown in Fig. 4A are obtained. These DFT calculations show strong surface approaching and adsorption selectivity for different molecular heading configurations. *We emphasize that we have also performed similar surface adsorption energy analysis using ion groups and heading configurations. The results lead to a similar conclusion of surface selectivity in a molecular diffusion model. The unit cells*

*and clusters analyses have shown even stronger surface-selectivity requirement. Therefore, for simplicity of presentation we only show calculations resulted from KDP molecules. The conclusion remains similar for other constituents.*

We further note that the theory of crystal nucleation by cluster groups (*4,8*) assumes that the surface crystallization is achieved dominantly by adsorption of nearly spherical clusters. However, this theory also argues that these nearly spherical clusters with critical radii favorable for crystallization *must have a similar lattice structure as crystalline surface*. Thus, even the clusters may have nearly spherical physical shapes the similar lattice structure requirement necessarily implies that they must have stacked planes of same orientation. In addition, they must have the molecular structures that are similar to the crystalline surface for which it will be bond with. This again favors surface selectivity.

**V. Highly-efficient stimulated Raman generation using TPPCl single-crystalline fiber produced by growth acceleration technique**

Multiple-phenyl phosphorous compounds [for which tetraphenyl-phosphonium-X family (TPP-X with X being Cl, Br, I, etc.) is an example] are a group of chemical materials that have been used as catalysts in variety of chemical processes and organic synthetic reactions ranging from paint production to pharmaceutical ingredients/intermediates manufacturing. Therefore, the demonstration of self-channeling-stimulated acceleration growth technique using some of these organic compounds has significant ramifications in applications. We choose TPP-X family single-crystals to illustrate the broad viability of the growth acceleration method both for their prolific chemical engineering and pharmaceutical applications as well as for immediate applications in advanced chip-size photonics device engineering due to their very large nonlinear optical coefficients.

We first note that single-pass CW Raman frequency conversion is technically very challenging. A recent comprehensive review (*11*) on all viable Raman frequency conversion schemes and techniques has shown that all current technologies require cavity enhancement and the typical Raman threshold is above 1 W. Furthermore, a large fraction of these schemes operate at high-peak-power pulsed mode with bulky high energy pulsed pump laser.

To generate CW Raman radiation we use a TPPCl fiber with unpolished facets as a Raman medium. Typical fiber size is 15 to 17 μm in diameter and 3 to 4 mm in length. The CW 532 nm pump laser has a power less than 4 mW. Up to ~ 64 nW CW Raman radiation (not counting the loss by the exit facet) was detected, indicating a Raman power conversion coefficient $> 3 \times 10^{-4}$. The loss due to the measurement optics and the single-crystal fiber facets is analyzed using transmission spectroscopy using a Raman spectrometer. The transmittance of laser injection and collecting optics are measured to be 76.6% and 87.1%, respectively. Transmission measurements yield an estimate of 70% loss per fiber facet, agrees with the SEM characterization of the TPPCl fiber facets.

Figures S2A and S2B show the stimulated Raman scattering and shifting spectrum, as well as the measurement of Raman threshold at 581.2 nm. The onset of stimulated Raman generation shown in Fig. S2B was measured as ~ 225 μW using a 532-nm pump. This is more than three orders of magnitude lower than the Raman threshold of the best cavity-based Raman converter, pulsed or CW, and far lower than any single-pass Raman converted at such low excitation level and gain medium volume. In fact, even with cavity enhancement the current best CW Raman converters still require more than 1 W pump power to reach the Raman generation threshold. Thus, the mm-size Raman converter shown herein can serve as an ideal Raman seed-laser for current Raman converter schemes and techniques. We note that the stimulated Raman growth characteristics shown in Fig. S2B agrees well with studies reported in literature (for instance, Ref. 12 shows a 1.25 W Raman threshold followed by slower growth of Raman radiation). Fitting experimental data with stimulated Raman generation theory (*12,13*) yields a Raman gain constant of $g_1 \approx 0.75$ cm/W. This is an extremely large Raman gain constants, enabled by the very large nonlinear susceptibility of organic nonlinear single-crystals such as TPP-X family single-crystals and tight spatial confinement of optical field.

**Fig. S1.**

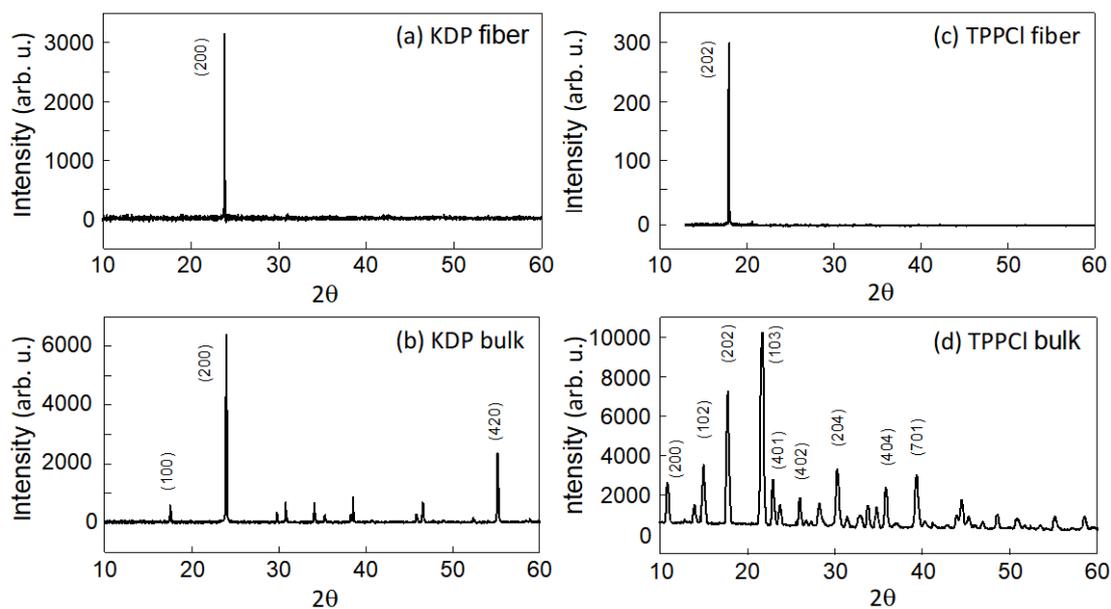

**XRD spectra of single crystal KDP (A, B) and TPPCl (C, D).**
Extremely sharp XRD peaks in **A** and **C** from single-crystalline fibers show no spectra broadening from defects and dislocations, excluding the presence of these factors.

**Fig. S2.**

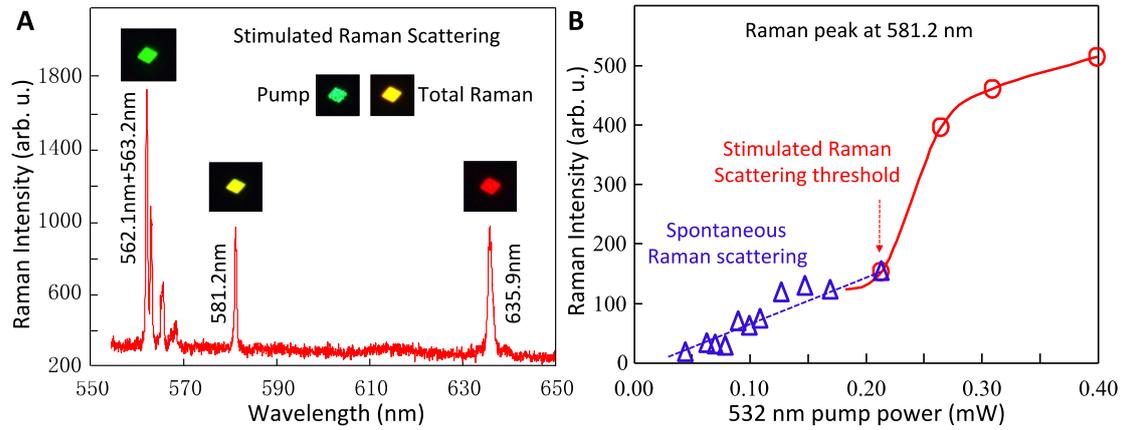

**Efficient stimulated Raman generation and frequency shift using a pump at 532 nm and a mm-size TPPCl fiber obtained by stimulated growth acceleration technique.**

(**A**) Stimulated Raman scattering spectra. (**B**) Raman threshold measurement.

# References

1. Y. Ren *et al.*, Growth of TPPB single crystal fibers, (to be published).
2. J. N. Homenuke, Numerical Solution of a Two-Component Reaction-Diffusion Equation in Two Spatial Dimensions. B.S. Thesis (2006).
   <http://laplace.physics.ubc.ca/People/jhomenuk/Thesis-JohnHomenuke.pdf>
3. P. Rudolph, Transport phenomena of crystal growth—heat and mass transfer. *AIP Conf. Proc.* **1270**, 107-132 (2010).
4. M. Bohenek, A. S. Myerson, W. M. Sun, Thermodynamics, cluster formation and crystal growth in highly supersaturated solutions of KDP, ADP and TGS. *J. Cryst. Growth* **179**, 213-225 (1997).
5. R. C. Prince, R. R. Frontiera, E. O. Potma, Stimulated Raman Scattering: From Bulk to Nano. *Chem. Rev.* **117**, 5070–5094 (2017).
6. D. W. H. Rankin, CRC handbook of chemistry and physics, 89th edition, edited by David R. Lide. *Crystallogr. Rev.* **15**, 223-224 (2009).
7. R. A. Laudise, *The growth of single crystals*. (Prentice Hall, 1970).
8. S. Kadam, Monitoring and Characterization of Crystal Nucleation and Growth during Batch Crystallization. Ph.D. Thesis (2012).
   <https://pdfs.semanticscholar.org/b633/a817141988e19d0d39afbda8a299127c66ee.pdf>
9. N. Zaitseva, L. Carman, Rapid growth of KDP-type crystals. *Prog. Cryst. Growth Charact. Mater.* **43**, 1-118 (2001).
10. P. Giannozzi *et al.*, QUANTUM ESPRESSO: a modular and open-source software project for quantum simulations of materials. *J. Phys.: Condens. Matter* **21**, 395502 (2009).
11. Q. Sheng, H. Ma, R. Li, M. Wang, W. Shi, J. Yao, Recent progress on narrow-linewidth crystalline bulk Raman lasers. *Results Phys.* **17**, 103073 (2020).
12. A. A. Demidovich, A. S. Grabtchikov, V. A. Lisinetskii, V. N. Burakevich, V. A. Orlovich, W. Kiefer, Continuous-wave Raman generation in a diode-pumped $Nd^{3+}$:$KGd(WO_4)_2$ laser. *Opt. Lett.* **30**, 1701–1703 (2005).
13. P. Černý, P. G. Zverev, H. Jelínková, T. T. Basiev, Efficient Raman shifting of picosecond pulses using BaWO4 crystal. *Opt. Commun.* **177**, 397–404 (2000).